%% file: o1_allsky3v1.tex
\documentclass[aps,prl,twocolumn,amsmath,amssymb,amsfonts,nofootinbib,superscriptaddress,altaffilletter]{revtex4-1}
\usepackage{graphicx}
\usepackage{hyperref}
\hypersetup{
  bookmarksopen=true
}
\usepackage{fontenc}
\usepackage{amssymb}
\usepackage{amsmath}
\usepackage{color}
\usepackage{supertabular}
\usepackage{dcolumn}
\usepackage[normalem]{ulem}

\def\sci#1#2{#1\times10^{#2}}

\def\RAJ{\textrm{RA}_{\textrm J2000}}
\def\DECJ{\textrm{DEC}_{\textrm J2000}}

\begin{document}

\title{ 
Sensitivity improvements in the search for periodic gravitational waves using O1 LIGO data
}

\author{Vladimir Dergachev}
\email{vladimir.dergachev@aei.mpg.de}
\affiliation{Max Planck Institute for Gravitational Physics (Albert Einstein Institute), Callinstrasse 38, 30167 Hannover, Germany}
\affiliation{Leibniz Universit\"at Hannover, D-30167 Hannover, Germany}

\author{Maria Alessandra Papa}
\email{maria.alessandra.papa@aei.mpg.de}
\affiliation{Max Planck Institute for Gravitational Physics (Albert Einstein Institute), Callinstrasse 38, 30167 Hannover, Germany}
\affiliation{Leibniz Universit\"at Hannover, D-30167 Hannover, Germany}
\affiliation{University of Wisconsin Milwaukee, 3135 N Maryland Ave, Milwaukee, WI 53211, USA}

\begin{abstract}
We demonstrate a breakthrough in the capabilities of robust, broad-parameter space searches for continuous gravitational waves. With a large scale search for continuous gravitational waves on the O1 LIGO data, we prove that our Falcon search achieves the sensitivity improvements expected from the use of a long coherence length, while maintaining the computational expense within manageable bounds. On this data we set the most constraining upper limits in the gravitational wave amplitude in the band 100-200 Hz. We provide full outlier lists and upper limits near 0-spindown band suitable for analysis of signals with small spindown such as boson condensates around black holes.
\end{abstract}

\maketitle


The term ``continuous gravitational waves'' refers to persistent nearly monochromatic gravitational waves, whose frequency $f$ changes slowly with time (typically $\dot{f}<10^{-12} f^2$). These signals are expected from rapidly rotating compact stars such as neutron stars due to a variety of mechanisms \cite{Lasky:2015uia}, as well as from more exotic scenarios \cite{Brito:2017zvb,boson1,boson2,boson3,Horowitz:2019aim,Horowitz:2019pru}. The resulting signal shape is largely independent of the specific emission channel. 

Continuous gravitational waves have not yet been detected and may well be the next big discovery of gravitational wave astronomy, unveiling the invisible population of galactic compact stars. The challenge is that when the waveform parameters are not known through some other observation channel, broad surveys become necessary.  These are computationally limited and the sensitivity and breadth actually attainable depend on the speed with which a search can be performed. The science potential of these searches depends crucially on their sensitivity/computing performance. 

The most sensitive and time-consuming searches are the completely coherent ones, where all available data is combined coherently and matched against the signal waveform. The smallest detectable signal amplitude for these searches decreases with the square root of the quantity of data, so the longer the data set is, the higher is the sensitivity of the search. On the other hand the resolution of different signal-waveforms grows with a higher power (say, 5) of the observation time and with it the computational cost. In short the best sensitivity, i.e. with fully coherent searches, is gained at very high computational expense. 

Semi-coherent searches allow to survey a broad range of different waveforms with a limited computational budget. They are called semi-coherent because the data is partitioned in shorter duration segments, each segment is searched coherently and the results from all segments combined (incoherently). Sensitivity is lost because each segment comprises less data than the whole data set, but much computational expense is saved because of the shorter time-baseline of the segments compared to the whole observation time.

Different semi-coherent search methods have been developed, that perform best in different conditions, for instance in very broad surveys, in very deep surveys, in very disturbed data, if the signal does not completely follow the model, for fast turn-around of results and more. All these are ``tuning parameters" of sort, that give rise to rather different search procedures \cite{Walsh:2016hyc,depths}.

Among the most sensitive methods -- if not the most sensitive for exploring a very broad range of waveforms and maintaining  robustness to deviations of the signal waveform from the exact model waveform -- is the so-called Powerflux search. Powerflux is one of the historical continuous wave search methods, with its first implementation dating back more than a decade \cite{Abbott:2007td}. Over the years and thanks to the experience accrued by exercising each new development on real data, the method has evolved significantly, becoming more sensitive and more resilient to all sorts of noise artefacts \cite{Abbott:2007td,Abbott:2008rg,Abadie:2011wj,Aasi:2015ssf,Abbott:2016udd,Abbott:2017mnu,Abbott:2018bwn,Dergachev:2019pgs}. 

This paper is a demonstration of the breakthrough performance of the last-generation Powerflux search, the Falcon ({\bf{Fa}}st {\bf{L}}oosely {\bf{Co}}here{\bf{n}}t).
The core of Falcon is a {\it {very efficient}} implementation of the so-called  {\it{loosely coherent}} approach, which enables searches over large parameter spaces with coherent timescales significantly longer than previously possible. Additionally the computational expense of these longer coherent time baseline searches remains manageable. This allows a sensitivity improvement with over two orders of magnitude gains in computational efficiency, while retaining the robustness of the Powerflux approach.

The original PowerFlux computes a sum of powers from the Fourier transforms of overlapping data segments from multiple detectors. Because of the overlap, some coherence between neighbouring segments is enforced and this can be considered as a very simple loosely coherent method. With this simple approach, however, in order to keep the signal power concentrated in frequency, the maximum length of the segments is limited by the requirement that the instantaneous frequency of the signal does not shift by more than a frequency bin during the segment duration.

%
Loosely coherent methods \cite{loosely_coherent, loosely_coherent2, loosely_coherent3} were developed initially to perform follow-ups of PowerFlux outliers with larger coherence lengths, while simultaneously retaining robustness to deviations of the signal from the assumed model. They use 
bilinear functions of powers from the Fourier transforms of overlapping data segments.

There is a simple conceptual reason for the performance attained by Falcon through the use of loosely-coherent methods. In general we can view semi-coherent methods as methods that are not only sensitive to physical signals, but also to a large set of unphysical signals. The unphysical signals are all the ones with phase jumps at segment boundaries, which the semi-coherent methods are insensitive to. The sensitivity loss of semi-coherent methods with respect to fully coherent ones can then be viewed as due to a huge trials factor resulting from this set of unphysical signals. Loosely coherent methods control the set of detectable signals and do this in ways that can provide additional benefits:  i) the bilinear kernel can be constructed to exclude unphysical phase jumps while allowing smooth deviations of the signal from the standard waveforms, for example Falcon can detect signals having extra phase modulations due to the presence of large planets or stars \cite{Singh:2019han} ii) the kernels can also be designed to efficiently approximate a fully coherent search over some time-scale longer than the segments iii) computational efficiency can be improved by engineering the set of detectable signals to suit computing hardware. These methods are quite complex in their specific implementation and we refer the interested reader to \cite{loosely_coherent} for more details.

We describe now a full-blown all-sky search using Falcon with a time baseline of 4\,hr, which is four times longer than ever used in similar searches (see for example \cite{Abbott:2017mnu,Abbott:2018bwn}). Since any candidate that does not pass the threshold on the first stage cannot be recuperated in later stages, the sensitivity of the first stage largely determines the overall sensitivity of the search. So the ability to perform an all-sky long coherence-length {\it{first-stage}} search marks an important breakthrough in scalability of continuous gravitational wave search methods. The purpose of this paper is to demonstrate such ability. In doing so, we also perform the most sensitive search on O1 data between 100 and 200 Hz.

In order to make the comparison simplest, we use the same data and very similar waveform parameter space as in \cite{Abbott:2017mnu}: We use data from LIGO's first observation run (O1) \cite{aligo, o1_data, losc} and search for continuous gravitational waves with frequencies between 20-200\,Hz  and frequency derivative (spindown) from $\sci{-1}{-8}$ through $\sci{1.11}{-9}$\,Hz/s. The computing expense of the first stage of this search is about 30 thousand core-days which, could be comfortably processed in less than a week on the busy Atlas cluster \cite{Atlas}. For comparison, the first stage of \cite{Abbott:2017mnu}, with a coherence length four times shorter than the one used in this search, used about 13 thousand core-days.

\begin{table}[htbp]
\begin{tabular}{c|c|c}
\hline
\hline
Stage & Coherence & SNR \\
	 &  length (hrs) & threshold\\
\hline
\hline
0  & 4 & 8 \\
1  & 4 & 8 \\
2  & 8 & 9 \\
3  & 24 & 10.5 \\
4  & 72 & 17 \\
\hline
\hline
\end{tabular}
\caption{Parameters of 5-stage pipeline used for analysis. Stage-1 uses finer grid spacings than Stage-0 in order to decrease the uncertainty on the signal parameter values and the computational expense of the Stage-2 follow-up. }
\label{tab:pipeline_parameters}
\end{table}

For this analysis we start with a coherence length of 4 hours and continue with 4 stages of follow-up searches on the candidates that survive each of the preceding stages. The coherent time-baseline and the signal-to-noise (SNR) threshold for each stage are given in Table \ref{tab:pipeline_parameters}. The last stage has a 3-day coherence length and is also carried out on each interferometer data separately. The multi-interferometer result and the single-interferometer results must display consistent parameters, as well as pass the respective detection thresholds for a candidate to pass Stage-4.

The list of candidates that survive Stage-4 is available in \cite{data}. Table \ref{tab:Outliers} shows a summary of that list, by taking the largest-SNR candidate from 0.1\,Hz bands. For convenience the summary excludes all candidates within 0.01\,Hz of multiples of 0.5\,Hz, which are induced by 0.25\,Hz combs of instrumental lines \cite{O1LowFreq}. The top 5 outliers are caused by hardware-simulated signals (the complete list is available at \cite{cw-o1-injection-params}) that are in the data stream to enable consistent signal recovery diagnostics for all search pipelines. 
The rest of the outliers are close to evident noise disturbances. The same holds true for all outliers in \cite{data}.

We provide 95\% confidence level upper limits on the gravitational wave amplitude, using the same procedure \cite{universal_statistics} as in many previous searches (see for example \cite{Abbott:2017mnu,Abbott:2018bwn}). 
We briefly outline the procedure: we derive upper limits across the entire parameter-space; 
for every 0.125 Hz band we then take the highest upper-limit with respect to polarization, sky-position, spindown and frequency. We note that the result of the maximization over all polarizations is dominated by linearly polarized waveforms. 
We refer to this upper limit as the worst-case upper limit (worst-case with respect to the source polarization/orientation). It provides strictly valid upper limits at the quoted confidence level for any waveform in the target parameter space. Additionally we compute upper limits by fixing the polarization to be circular and by maximizing over all the other parameters. This is informative of the minimum detectable signal amplitude when the source orientation/polarization parameters are optimal for detection, and hence yields the maximum reach of the search. We recall that the effective gravitational wave amplitude scales proportionally to $1=\cos^2\iota$ and to $\cos\iota$ for the + and x polarizations respectively. When $\cos\iota=0$ the line of sight of the observer is perpendicular to the rotation axis of the star, and from a distorted neutron star the radiation that observer sees is linear, whereas when $\cos\iota=\pm 1$ the view is along the rotation axis and the radiation circularly polarized.

The upper limits set with other procedures, for instance \cite{Abbott:2017mnu,Abbott:2018bwn,EHO1,Pisarski:2019vxw,Ming:2019xse}, have a slightly different meaning. They represent the gravitational wave intrinsic amplitude that the target population of signals has to have so that a fraction of it, equal to the confidence level, would be detected by the search. They are a sort of population-average upper limits. Unlike the PowerFlux upper limits that hold true for any signal waveform, there exist small portions of the waveform parameter space for which the \cite{Abbott:2017mnu,Abbott:2018bwn,EHO1,Pisarski:2019vxw,Ming:2019xse} upper limits do not hold.

To ease comparison with these other upper limits we introduce a proxy for the population-average upper limits for this search as the weighted average of upper limits from individual polarizations. The weights were heuristically determined.
To provide true population-average upper limits in every frequency band one must perform extensive fake-signal simulation-and-recovery studies. This is not practical as the computational expense would be very considerable. Instead we verify our proxy by adding 1440 signals with frequency uniformly distributed in the 20-200\,Hz range to the real, and measuring how effectively our pipeline recovers them. The uniform distribution is chosen to sample the full variety of detector artefacts. Sky location and orientation of the source are also chosen uniformly. The spindown is log-uniform distributed between $\sci{-7.5}{-10}$ and $\sci{-2.3}{-11}$\,Hz/s. 

For each simulated signal the analysis is carried out just as it in the real search, including the follow-up stages. A signal is considered detected if there is an outlier in the final outlier table within 50\,$\mu$Hz of the true injection frequency $f_0$, within $\sci{2.5}{-11}$\,Hz/s of true injection spindown and within $1.5\,{\textrm{Hz}}/f_0$ in ecliptic distance (distance between projection of outlier and injection location from unit sphere onto the ecliptic plane). 

Above 50\,Hz at the $h_0$ value of the proxy upper limit the detection rate is 95\%, while below 50\,Hz the rate is 80\%.  The lower detection rate at low frequencies is due to heavy contamination with detector artifacts.

\begin{figure}[htbp]
\begin{center}
  \includegraphics[width=3.3in]{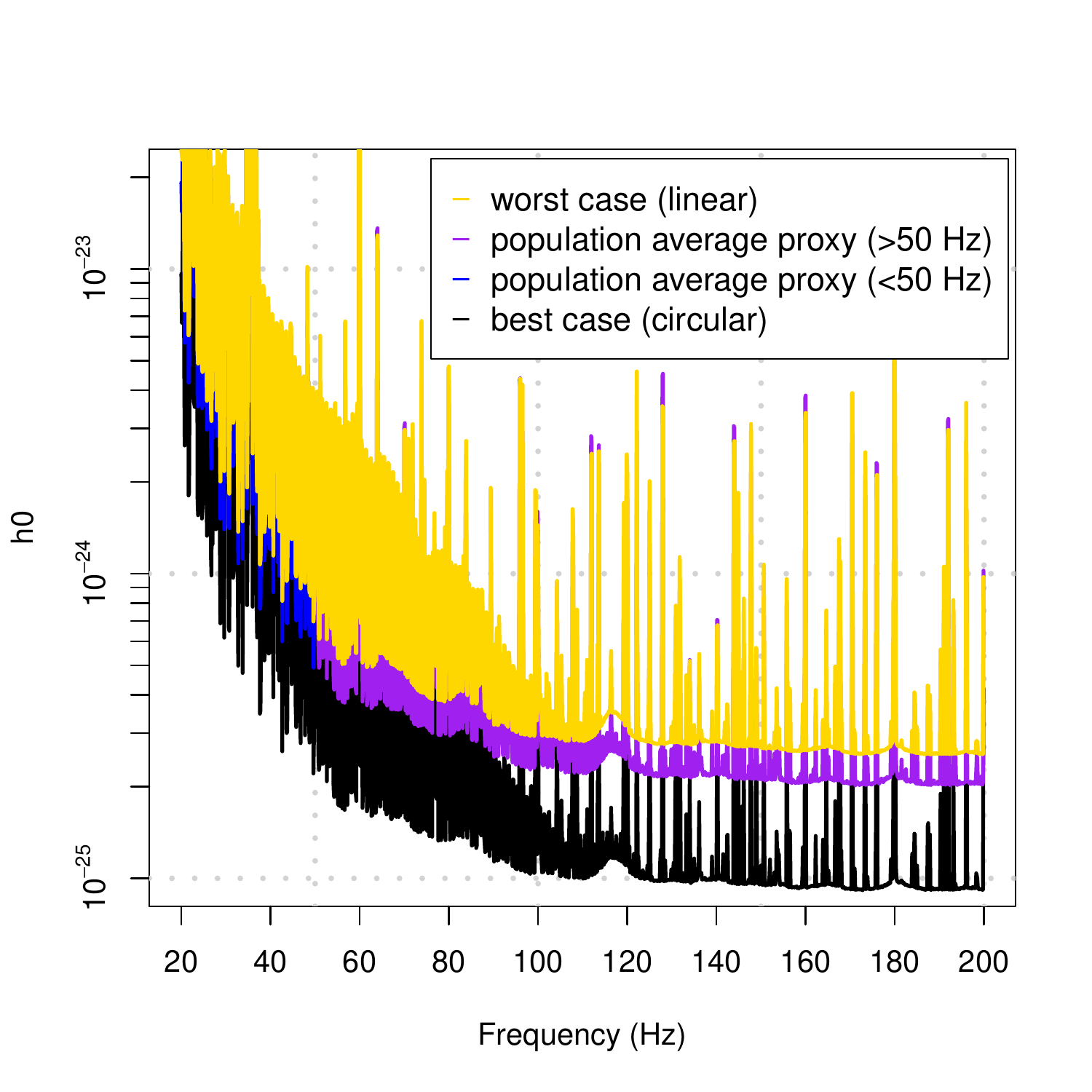}
 \caption{Upper limits on the intrinsic gravitational wave amplitude at the detectors plotted against signal frequency. The top curve (yellow) shows worst-case upper limits, the next lower curve (purple and blue) the population-average proxy, followed by black curve showing the circular polarization upper limits. The different colours of population average proxy mark regions with 80\% recovery rate ($<50$\,Hz) and 95\% recovery rate ($>50$\,Hz), as explained in the text.}
\label{fig:O1_upper_limits}
\end{center}
\end{figure}

Figure \ref{fig:O1_upper_limits} shows the upper limit values as a function of signal frequency.  As already well known \cite{EHO1, covas, Abbott:2017mnu}, the frequency range below 100 Hz in this data set is plagued by many instrumental artefacts, including a nefastous 0.25\,Hz comb of  lines, which significantly limit the sensitivity of any search for signals with significant energy content in this region. This manifests itself in our results with many upper limit values being significantly increased in this frequency range, with respect to the value they would have in the absence of coherent disturbances. This is what is producing a much larger variance in the distribution of upper limit values. The ``floor" of the curves in Figure \ref{fig:O1_upper_limits} remains however representative of the performance of the search in most of the frequency bands at higher frequencies. Following \cite{depths}, we compute the sensitivity depth corresponding to our upper limits as square root of power spectral density divided by the upper limit values and we report it in Table \ref{tab:depths}. 

The sensitivity gain of this search with respect to the previously established PowerFlux upper limits \cite{O1LowFreq}
is about 30\%. This is consistent with the increase in coherence length by a factor of 4 in the first stage of the search over the O1 data set. Although 30\% might not appear to the non-expert to be a very large sensitivity improvement, in the context of continuous gravitational wave searches it is: it translates into an increase of nearly 3 in volume of space visible from Earth and it corresponds to a larger increase in sensitivity than that achieved with many months of commissioning work at the LIGO detectors between the first and second science runs. 


\begin{table}[h]
\begin{center}
\begin{tabular}{rccc}\hline
 & \multicolumn{1}{c}{Worst-case} & \multicolumn{1}{c}{Best-case} & \multicolumn{1}{c}{Population average}\\
\hline
\hline
Depth [Hz$^{-1/2}$] & 24.5 & 70.0 & 30.7\\
Depth error [Hz$^{-1/2}$] & \,\,\,6.5 & 16.5 & \,\,\,8.4 \\
\hline
\end{tabular}
\end{center}
\caption{Average sensitivity depth from our upper limits from this search. Following \cite{depths} we also provide the standard deviation on the average as ``depth error''. The large values of the depth error are due to the numerous combs of instrumental lines. }
\label{tab:depths}
\end{table}

\begin{figure}[htbp]
\includegraphics[width=3.3in]{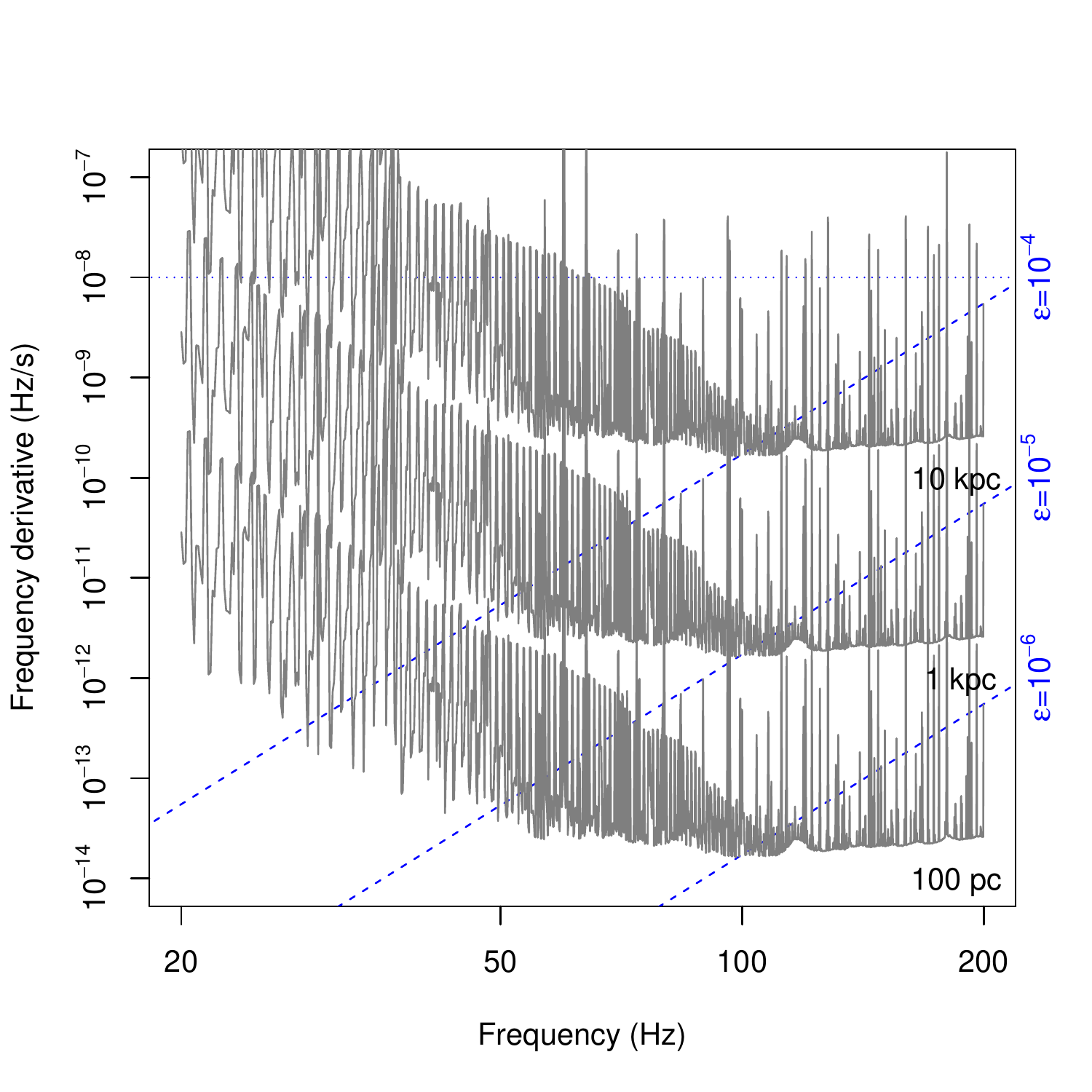}
\caption[Spindown range]{
\label{fig:spindown_range}
Range of the search for neutron stars
spinning down solely due to gravitational radiation.  This is a
superposition of two contour plots.  The grey solid lines are contours of the maximum distance at which a neutron
star could be detected as a function of gravitational wave frequency
$f$ and its derivative $\dot{f}$.  The dashed lines 
are contours of the corresponding ellipticity
$\epsilon(f,\dot{f})$. The fine dotted line marks the maximum spindown searched. Together these quantities tell us the
maximum range of the search in terms of various populations.}
\end{figure}

We translate the circular polarization (best-case) upper limits on the intrinsic gravitational wave amplitiude $h_0$ in maximum reach of the search for signals with a given frequency and spindown. This maximum reach corresponds to the situation when all the rotational energy lost is carried away by gravitational waves. This is shown in Figure \ref{fig:spindown_range}. We assume an equatorial quadrupolar ellipticity of the star to be the cause of the continuous gravitational wave emission and we show the iso-ellipticity curves on the same plot. At the high end of the frequency range this search is sensitive to a source with $10^{-6}$ equatorial ellipticity up to 440\,pc away. It is known that neutron stars can readily support equatorial ellipticities of more than $10^{-6}$ \cite{crust_limit, crust_limit2}.

Boson condensates around black holes are another potential source of continuous gravitational waves \cite{boson1, boson2, boson3}. These signals are expected to have a non appreciable spindown, so we can examine the results of this search for zero spindown waveforms and determine its sensitivity for boson condensate signals around isolated galactic black holes. The upper limit data \cite{data} includes a separate set of upper limits covering near 0 frequency derivatives. As an example, we include a figure showing the detection reach for vector boson condensate with parameter $\alpha=0.03$ around black holes with spin $0.2$. Using the data provided in \cite{data} the interested reader can derive the upper limits corresponding to any choice of $\alpha$ and spin.
\begin{figure}[htbp]
\includegraphics[width=3.3in]{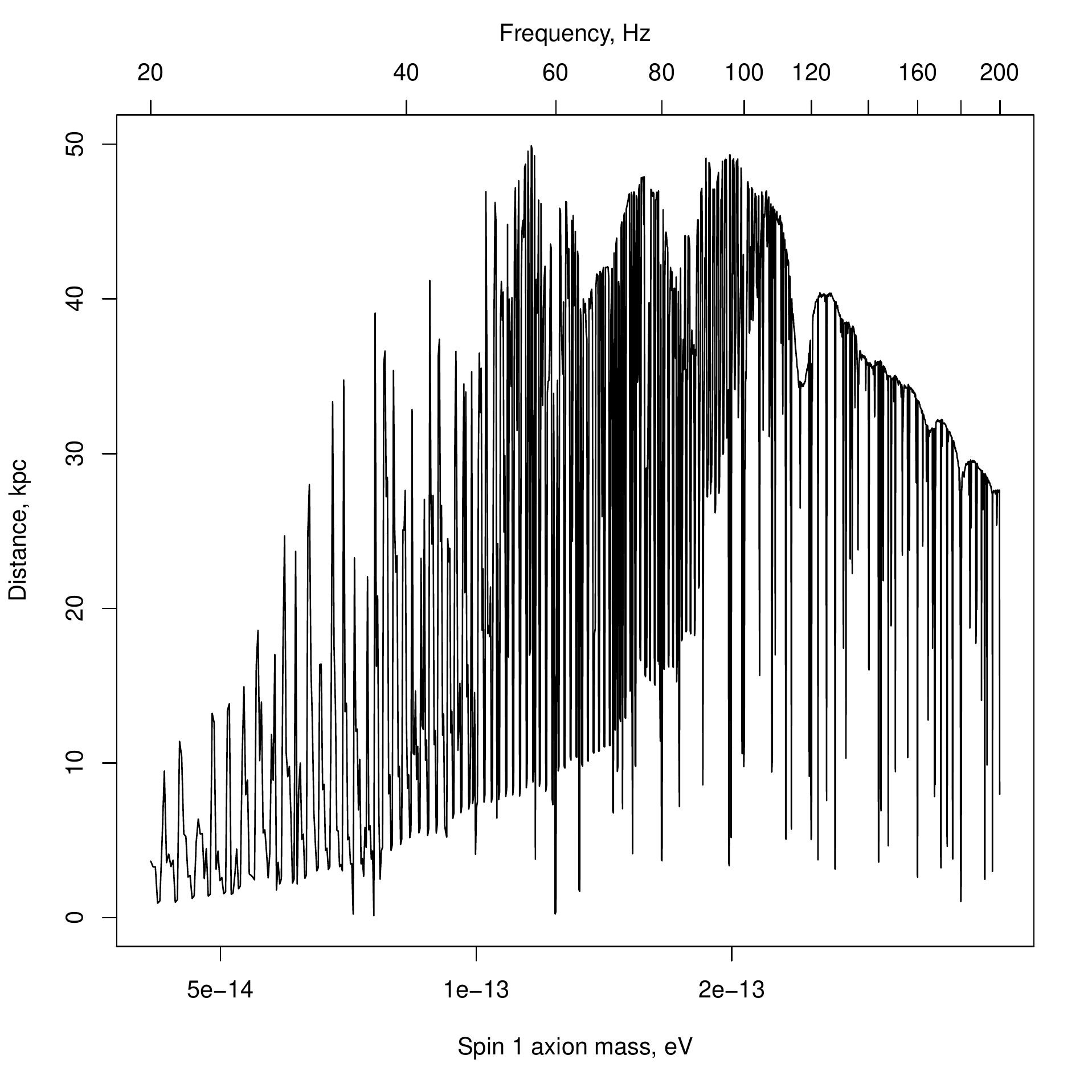}
\caption[Spindown range]{
\label{fig:axion_range}
Sensitivity range to signals from vector boson condensates with parameter $\alpha=0.03$ around black holes with spin $0.2$. This plot was produced using worst-case near-0 spindown upper limits.}
\end{figure}

\begin{table*}[htbp]
\begin{center}
\scriptsize
\begin{tabular}{D{.}{.}{2}D{.}{.}{3}D{.}{.}{5}D{.}{.}{4}D{.}{.}{4}D{.}{.}{4}l}\hline
\multicolumn{1}{c}{Idx} & \multicolumn{1}{c}{SNR}   &  \multicolumn{1}{c}{Frequency} & \multicolumn{1}{c}{Spindown} &  \multicolumn{1}{c}{$\RAJ$}  & \multicolumn{1}{c}{$\DECJ$} & Description \\
\multicolumn{1}{c}{}	&  \multicolumn{1}{c}{}	&  \multicolumn{1}{c}{Hz}	&  \multicolumn{1}{c}{nHz/s} & \multicolumn{1}{c}{degrees} & \multicolumn{1}{c}{degrees} & \\
\hline \hline
\input{outliers.table}
\hline
\end{tabular}
\caption[Outliers that passed detection pipeline]{Outliers that passed detection pipeline excluding outliers within 0.01\,Hz from 0.25 Hz comb of instrumental lines. Only the highest-SNR outlier is shown for each 0.1\,Hz frequency region. Outliers marked with ``line'' had strong narrowband disturbances identified near the outlier location.
Frequencies are converted to epoch GPS $1130529362$.} 
\label{tab:Outliers}
\end{center}
\end{table*}

Powerflux is one of the most effective methods to search for a very broad set of continuous gravitational wave signals over month-long data sets. Its main sensitivity limit comes from the constrain that the initial coherence length cannot exceed the length of time during which the frequency of a signal might move by more than a frequency bin; this feature is common to all other semi-coherent searches, see for example Table II of \cite{Abbott:2017mnu}. The Falcon search overcomes this constraint. The optimised algorithmic implementation heavily uses vector processing units and dynamic programming \cite{loosely_coherent} and achieves improvements in computational efficiency of hundreds. Such improvements make it possible to actually perform a search with a significantly longer coherence length than previously achieved on an in-house computing cluster, in the first and broadest parameter space stage of the search. This fact is extremely important because the sensitivity of the first stage largely determines the overall sensitivity. With a 5-stage hierarchical procedure on highly disturbed and hence unpredictable data, the search presented here demonstrates that the Falcon search, can be successfully deployed in real searches, that the computational expense does not increase substantially while the sensitivity improvements are consistent with those expected from the increased initial coherence length. This sets the stage for a new era of semi-coherent robust continuous wave searches on the next sets of gravitational wave data.

The search was performed on the ATLAS cluster at AEI Hannover, for which we thank Bruce Allen. We also thank Carsten Aulbert and Henning Fehrmann for their support.

This research has made use of data, software and/or web tools obtained from the LIGO Open Science Center (\url{https://losc.ligo.org}), a service of LIGO Laboratory, the LIGO Scientific Collaboration and the Virgo Collaboration. LIGO is funded by the U.S. National Science Foundation. Virgo is funded by the French Centre National de Recherche Scientifique (CNRS), the Italian Istituto Nazionale della Fisica Nucleare (INFN) and the Dutch Nikhef, with contributions by Polish and Hungarian institutes.

\end{document}

%% file: outliers.table
1 & 870 & 52.80832 & 0.006 & 306.634 & -83.997 & Hardware injection ip5\\
2 & 637 & 191.03126 & -8.652 & 351.425 & -33.552 & Hardware injection ip8\\
4 & 387 & 146.16934 & -6.710 & 359.608 & -65.199 & Hardware injection ip6\\
5 & 376 & 38.47793 & -6.235 & 332.323 & -14.679 & Hardware injection ip12\\
6 & 300 & 108.85718 & -0.006 & 178.641 & -33.400 & Hardware injection ip3\\
22 & 54 & 99.97667 & -5.115 & 100.314 & -41.321 & coincident contamination in LHO and LLO\\
23 & 50 & 31.51238 & -5.619 & 226.702 & -23.180 & Heavy contamination, 0.25 Hz comb in H1, 31.512 Hz line in L1\\
33 & 27 & 65.51035 & -5.419 & 198.120 & -40.763 & 0.25 Hz comb of instrumental lines\\
36 & 23 & 32.69785 & -9.940 & 45.757 & -37.300 & Lack of coherence, contamination in H1\\
38 & 21 & 82.51584 & -3.994 & 157.388 & -46.681 & Lack of coherence, 0.25 Hz comb in H1\\
39 & 21 & 81.52983 & -6.310 & 332.731 & -45.648 & 0.25 Hz comb of instrumental lines\\
40 & 21 & 107.13643 & -6.677 & 12.094 & -57.316 & coincident artifacts at 107.12 Hz in H1 and L1\\
42 & 20 & 113.01128 & 1.044 & 304.688 & 9.253 & 0.25 Hz comb of instrumental lines\\
45 & 19 & 90.65642 & -6.944 & 302.827 & 59.828 & no coherence, disturbed H1 spectrum\\
46 & 19 & 62.80672 & -7.985 & 276.491 & -12.515 & Sharp bin-centered line in L1 at 62.8 Hz\\
47 & 19 & 45.01809 & -1.227 & 186.589 & 28.604 & Lines in H1 and L1 at 45 Hz, contaminated spectrum\\
48 & 18 & 133.30755 & -6.548 & 124.281 & -50.348 & Line in L1 at 133.33 Hz\\
49 & 18 & 49.96416 & -9.906 & 335.191 & -18.085 & Highly contaminated spectrum\\
50 & 18 & 164.68348 & -4.927 & 48.010 & -9.672 & Line in L1 at 164.7 Hz\\
51 & 18 & 86.51503 & -5.252 & 56.836 & -36.012 & Coincident lines at 86.5 Hz, 0.25 Hz comb, sloping spectrum in L1\\
52 & 18 & 48.98773 & -4.410 & 43.189 & -22.949 & Highly contaminated H1 and L1 spectrum near 49 Hz\\
53 & 18 & 192.83187 & -7.790 & 248.222 & 46.044 & Contaminated H1 spectrum\\
56 & 18 & 54.12124 & -9.790 & 225.480 & -16.962 & Sharp bin-centered line in L1 at 54.1 Hz\\
57 & 17 & 91.71446 & -8.506 & 128.301 & -28.413 & Disturbed spectrum in H1 and L1\\
58 & 17 & 53.93050 & -2.956 & 212.788 & -24.064 & Disturbed H1 spectrum\\
59 & 17 & 107.66022 & -9.660 & 24.618 & -6.139 & Sharp line in L1 at 107.7 Hz\\